\documentclass[12pt]{article}

\usepackage{epsfig}			%standard package. Note graphics<graphicx<epsfig 

\newcommand{\nb}[0]{{N_{B}}}
\newcommand{\ns}[0]{{N_{S}}}
\newcommand{\antin}[0]{\overline{n}}
\newcommand{\usualz}[0]{Z_{0, \nb -1}}
\newcommand{\sx}[0]{\sigma_x}
\newcommand{\sy}[0]{\sigma_y}
\newcommand{\sz}[0]{\sigma_z}
\newcommand{\cnotyes}[2]{\sx(#2)^{n(#1)}}
\newcommand{\cnotno}[2]{\sx(#2)^{\antin(#1)}}
\newcommand{\va}[0]{{\vec{a}}}
\newcommand{\vb}[0]{{\vec{b}}}

\newcommand{\ket}[1]{|#1\rangle}

\newcommand{\beq}{\begin{equation}}  
\newcommand{\eeq}{\end{equation}}
\newcommand{\beqa}{\begin{eqnarray*}}  
\newcommand{\eeqa}{\end{eqnarray*}}
\newcommand{\rarrow}[0]{\rightarrow}

\newcommand{\enote}[1]{\cite{#1}}    %call it \enote (endnote) rather than \cite
\newcommand{\eqlabel}[1]{\renewcommand{\theequation}{#1}}%this allows me to label 
\begin{document}
\title{A Rudimentary Quantum Compiler}

\author{Robert R. Tucci\\
        P.O. Box 226\\ 
        Bedford,  MA   01730\\
        tucci@ar-tiste.com}

\date{ \today} 

\maketitle

\vskip2cm
\section*{Abstract}
We present a new algorithm for reducing an arbitrary
unitary matrix into a sequence of elementary operations 
(operations such as controlled-nots and qubit rotations). Such a sequence of
operations can be used to manipulate an array of quantum bits (i.e.,
a quantum computer). We report on a C++ program 
called ``Qubiter" that implements our algorithm. Qubiter source code
is publicly available.

\newpage
\section*{1. Introduction}
\subsection*{1(a) Previous Work}
\mbox{}\indent	
In classical computation and digital electronics, one deals with sequences of elementary operations
(operations such as AND, OR and NOT).
These sequences are used to manipulate an array of classical bits.
The operations are elementary in the sense that they 
act on only a few bits (1 or 2) at a time.
 Henceforth, we will sometimes refer to sequences as products and to 
operations as operators, instructions, steps or gates.
 Furthermore, we will  abbreviate 
the phrase ``sequence of elementary operations" by ``SEO".
In quantum computation\enote{DiV}, one also deals with SEOs (with operations such as  
controlled-nots and qubit rotations), but for manipulating 
quantum bits (qubits) instead of classical bits.
SEOs are often represented graphically by a qubit circuit.

In quantum computation, one
often knows the unitary operator $U$ that describes the evolution 
of an array of qubits. One must then 
find a way to reduce $U$ into a
SEO. In this paper, we present a new algorithm 
for accomplishing this task. We also report on a C++ program called
``Qubiter" that implements our algorithm. Qubiter source code is publicly
available at www.ar-tiste.com/qubiter.html . We call Qubiter a ``quantum compiler" because, like a  
classical compiler, it produces a SEO for manipulating bits.

Our algorithm is general in the sense that it can be applied to any
unitary operator $U$. Certain $U$ are known to be polynomial in $\nb$;
i.e, they can be expressed as a SEO whose
number of gates varies as a polynomial in $\nb$ as $\nb$ varies, 
where $\nb$ is the number 
of bits. For example, a Discrete Fourier Transform (DFT) can be 
expressed as Order($\nb^2$) steps \enote{DFT}. Our algorithm 
expresses a DFT as Order($4^\nb$) steps.
Hence, our algorithm is not ``polynomially efficient"; i.e., it does
not give a polynomial number of steps for all $U$ for which this is possible.
However, we believe that it is possible to introduce optimizations into 
the algorithm so as to make it polynomially efficient. 
Future papers will report our progress in finding such optimizations.

Previous workers \enote{CastOfThousands} have described another 
algorithm for reducing a unitary operator into a SEO. 
Like ours, their algorithm is general, and it is not expected to
be polynomial efficient without optimizations. 
Our algorithm is significantly
different from theirs. Theirs is based on a mathematical technique
described in Refs.\enote{Murn}-\enote{Reck}, whereas ours is based on
a mathematical technique called CS Decomposition\enote{Stew}-\enote{PaiWei},
to be described later.

Quantum Bayesian (QB) Nets\enote{Tucci95}-\enote{QFog} are a method of modelling quantum
systems graphically in terms of network diagrams. In a companion paper\enote{Tucci98b}, we  show how to apply the
results of this paper to QB nets.

\subsection*{1(b) CS Decomposition}
\mbox{}\indent	
As mentioned earlier, our algorithm utilizes a mathematical technique
called CS Decomposition\enote{Stew}-\enote{PaiWei}. In this name, the letters C and S 
stand for ``cosine" and ``sine". Next we will state
the special case of the CS Decomposition Theorem
that arises in our algorithm. 

Suppose that $U$ is an $N\times N$ unitary matrix, where $N$ is an even number.
Then the CS Decomposition Theorem states that one can always 
express $U$ in the form 

\beq
U = 
\left [
\begin{array}{cc}
L_0 &  0 \\
0   &  L_1
\end{array}
\right ]
D
\left [
\begin{array}{cc}
R_0 &  0 \\
0   &  R_1
\end{array}
\right ]
\;,
\eqlabel{1.1}\eeq
where $L_0, L_1, R_0, R_1$ are 
$\frac{N}{2}\times \frac{N}{2}$  unitary matrices and

\beq
D = 
\left [
\begin{array}{cc}
D_{00} &  D_{01} \\
D_{10}  &  D_{11}
\end{array}
\right ]
\;,
\eqlabel{1.2a}\eeq

\beq
D_{00} = D_{11} = diag(C_1, C_2, \ldots, C_{\frac{N}{2}})
\;,
\eqlabel{1.2b}\eeq

\beq
D_{01} = diag(S_1, S_2, \ldots, S_{\frac{N}{2}})
\;,
\eqlabel{1.2c}\eeq

\beq
D_{10} = - D_{01}
\;.
\eqlabel{1.2d}\eeq
For $i \in \{ 1, 2, \ldots , \frac{N}{2} \}  $,
 $C_i$ and $S_i$
are real numbers that satisfy

\beq
C_i^2 + S_i^2 = 1
\;.
\eqlabel{1.2e}\eeq
Henceforth, we will use the term 
{\it D matrix} to refer to any matrix that satisfies Eqs.(1.2).
If one partitions $U$ into four blocks $U_{ij}$ 
of size $\frac{N}{2}\times \frac{N}{2}$ ,
then 

\beq
U_{ij} = L_i D_{ij} R_j
\;,
\eqlabel{1.3}\eeq
for $i, j \in \{0, 1\}$. Thus, $D_{ij}$ gives the singular values
of $U_{ij}$.

 More general versions of the CS Decomposition Theorem allow for the
possibility that we
partition $U$ into 4 blocks
of unequal size. 

Note that if $U$ were a general (not necessarily unitary) matrix, then
the four blocks $U_{ij}$ would be unrelated. 
Then to find the singular values of the four blocks $U_{ij}$ would require
 eight unitary matrices (two for each block), instead of the four $L_i, R_j$.
 This double use of the $L_i, R_j$ is a key property of the CS decomposition.

\subsection*{1(c) Bird's Eye View of Algorithm}
\mbox{}\indent	
Our algorithm is described in detail in subsequent sections.
Here we will only give a bird's eye view of it.

%don't float in general
%		\begin{figure}
		\begin{center}
			\epsfig{file=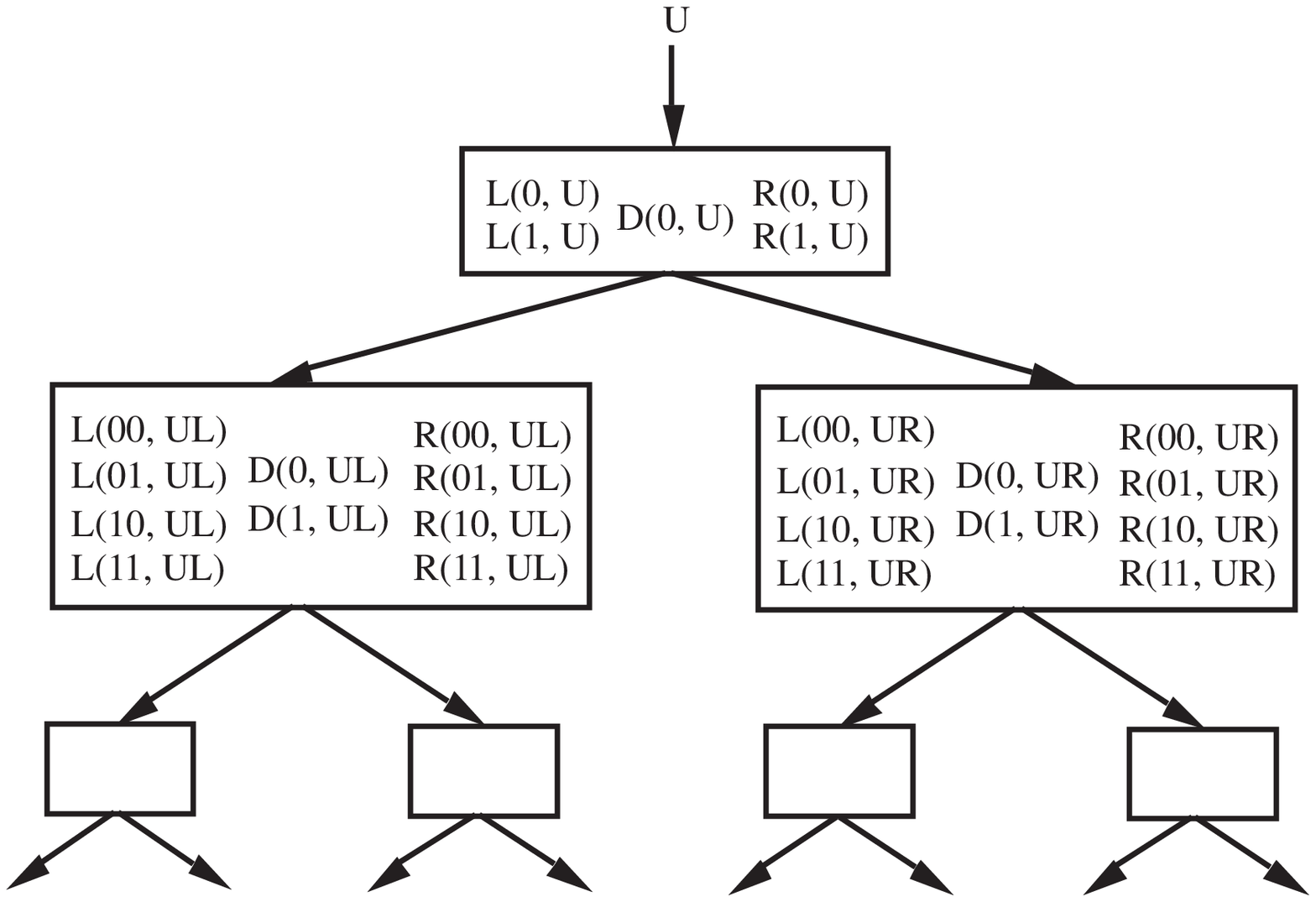}
			%\caption
			
			{Fig.1 A binary tree. Each node
			$\beta$ has a single parent. If the parent is to 
			$\beta$'s right (ditto, left), then $\beta$
			contains the names of the matrices produced by
			applying the CS Decomposition Theorem to the $L$
			matrices (ditto, 
			$R$ matrices) of $\beta$'s parent.}
			%\label{}
		\end{center}
%		\end{figure}

%new paragraph after figure

Consider Fig.1. 
We start with a unitary matrix $U$.
Without loss of generality, we can assume that 
the dimension of $U$ is $2^\nb$ for some $\nb\geq 1$.
(If initially $U$'s dimension is not a power of 2,
 we replace it by a direct sum
$U\oplus I_r$ whose dimension is a power of two.)
We apply the CS Decomposition method to $U$. This yields a
matrix $D(0, U)$ of singular values, two unitary matrices 
$L(0, U)$ and $L(1, U)$ on the left and two unitary
matrices $R(0, U)$ and $R(1, U)$ on the right.
Then we apply the CS Decomposition method to 
each of the 4 matrices 
$L(0, U), L(1, U), R(0,U)$ and $R(1, U)$ that were produced
in the previous step. Then we apply the CS Decomposition
method to each of the 16 $R$ and $L$ matrices that were produced
in the previous step. And so on. The lowest row of the pyramid 
in Fig.1 has
$L$'s and $R$'s that are $1\times 1$ dimensional, i.e., just complex numbers.

Call a {\it central matrix} either (1) a single D matrix, or 
(2) a direct sum $D_1 \oplus D_2 \oplus \cdots  \oplus D_r$ of D matrices,
or (3) a diagonal unitary matrix. From Fig.1 it is clear that 
the initial matrix $U$ can be expressed as a product of 
central matrices, with each node of the tree providing
one of the central matrices in the product. Later on we will 
present techniques
for decomposing any central matrix into a SEO.

\section*{2. Preliminaries}
\mbox{}\indent	
In this section, we introduce some notation and some general mathematical
concepts that will be used in subsequent sections.

\subsection*{2(a) General Notation}
\mbox{}\indent	
We define
$Z_{a, b} = \{ a, a+1, \ldots, b\}$ for any integers $a$ and $b$.
$\delta(x,y)$ equals one if $x=y$ and zero otherwise.

We will use the symbol $\nb$ for the number ($\geq 1$) of bits and 
$\ns = 2^\nb$ for the number of states with $\nb$ bits. 
Let $Bool = \{0,1\}$. We will use lower case Latin letters
$a,b,c\ldots \in Bool$ to represent bit values and
lower case Greek letters $\alpha, \beta, \gamma, \ldots \in \usualz $
to represent bit positions. A vector such as 
$\va = a_{\nb-1} \ldots a_2 a_1 a_0$ will represent 
a string of bit values, $a_\mu$ being the value of the $\mu$'th bit
for $\mu \in \usualz$. A bit string $\va$ has
a decimal representation 
$d(\va) = \sum^{\nb-1}_{\mu=0} 2^\mu a_\mu$.
For $\beta\in \usualz$, we will use 
$\vec{u}(\beta)$ to denote the $\beta$'th standard unit vector, i.e, 
the vector with bit value of 1 at bit position $\beta$ and
bit value of zero at all other bit positions.

$I_r$ will represent the $r$ dimensional unit matrix.
Suppose $\beta\in \usualz$ and $M$ is any $2\times 2$ matrix. We define
$M(\beta)$ by
\beq
M(\beta) = 
I_2 \otimes
\cdots \otimes 
I_2 \otimes 
M \otimes 
I_2 \otimes
\cdots \otimes 
I_2
\;,
\eqlabel{2.1}\eeq
where the matrix $M$ on the right side is located
at bit position $\beta$ in the tensor product of $\nb$ $2\times 2$ matrices. The numbers
that label bit positions in the tensor product increase from
right to left ($\leftarrow$), and the rightmost bit is taken
to be at position 0.

For any two square matrices $A$ and $B$ of the same dimension,
 we define the $\odot$ product by
 $A \odot B = A B A^{\dagger}$, where 
$A^{\dagger}$ is the Hermitian conjugate of $A$.

$\vec{\sigma} = (\sx, \sy , \sz)$ will 
represent the vector of Pauli matrices, where

\beq
\sx=
\left(
\begin{array}{cc}
0&1\\
1&0
\end{array}
\right)
\;,
\;\;
\sy=
\left(
\begin{array}{cc}
0&-i\\
i&0
\end{array}
\right)
\;,
\;\;
\sz=
\left(
\begin{array}{cc}
1&0\\
0&-1
\end{array}
\right)
\;.
\eqlabel{2.2}\eeq

\subsection*{2(b) Sylvester-Hadamard Matrices}
\mbox{}\indent	
The Sylvester-Hadamard matrices\enote{Had} $H_r$ are 
defined by:

\beq
H_1 = 
\begin{tabular}{r|rr}
         & {\tiny 0} & {\tiny 1} \\
\hline
{\tiny 0}& 1&  1\\
{\tiny 1}& 1& -1\\
\end{tabular}
\;,
\eqlabel{2.3a}\eeq

\beq
H_2 = 
\begin{tabular}{r|rrrr}
         & {\tiny 00} & {\tiny 01} & {\tiny 10} & {\tiny 11}\\
\hline
{\tiny 00}&  1&  1&  1&  1\\
{\tiny 01}&  1& -1&  1& -1\\
{\tiny 10}&  1&  1& -1& -1\\
{\tiny 11}&  1& -1& -1&  1\\
\end{tabular}
\;,
\eqlabel{2.3b}\eeq

\beq
H_{r+1} = H_1 \otimes H_r
\;,
\eqlabel{2.3c}\eeq
for any integer $r\geq 1$. In Eqs.(2.3), 
 we have labelled the rows and columns with binary numbers
 in increasing order, and $\otimes$
indicates a tensor product of matrices. 
From Eqs.(2.3), one can show that
the entry of $H_r$ at row $\va$ and column $\vb$ is given by

\beq
(H_r)_{\va, \vb} = (-1)^{\va\cdot \vb}
\;,
\eqlabel{2.4}\eeq
where $\va\cdot \vb = \sum^{r-1}_{\mu=0} a_\mu b_\mu$. It is easy to 
check that

\beq
H_r^T = H_r
\;,
\eqlabel{2.5}\eeq

\beq
H_r^2 = 2^r I_r
\;.
\eqlabel{2.6}\eeq
In other words, $H_r$ is a symmetric matrix, and the inverse of $H_r$ equals
$H_r$ divided by however many rows it has.

\subsection*{2(c) Permutations}
\mbox{}\indent	
Subsequent sections will use the following very basic facts about permutations.
For more details, see, for example, Ref.\enote{Perm}.

A {\it permutation} is a 1-1 onto map from a finite set $X$ onto itself.
The set of permutations on set $X$ is a group if group multiplication
is taken to be function composition. $S_n$, the {\it symmetric group
in $n$ letters}, is defined as the group of all permutations
on any set $X$ with $n$ elements.
If $X= Z_{1,n}$, then a permutation $G$ which maps $i\in X$ to $a_i\in X$
(where $i\neq j$ implies $a_i\neq a_j$) can be represented by a 
matrix with entries

\beq
(G)_{j, i} = \delta(a_j, i)
\;,
\eqlabel{2.7}\eeq
for all $i,j \in X$. Note that all entries in any given row or column 
equal zero except for one entry which equals one.
 An alternative notation for $G$ is 

\beq
G = 
\left (
\begin{array}{ccccc}
1   & 2   & 3   & \cdots & n   \\
a_1 & a_2 & a_3 & \cdots & a_n
\end{array}
\right )
\;.
\eqlabel{2.8}\eeq
The product of two symbols of the type shown in Eq.(2.8) 
is defined by function composition. For example,

\beq
\left (
\begin{array}{ccc}
a_1 & a_2 & a_3 \\
b_1 & b_2 & b_3 
\end{array}
\right )
\left (
\begin{array}{ccc}
  1 &   2 &   3 \\
a_1 & a_2 & a_3 
\end{array}
\right )
=
\left (
\begin{array}{ccc}
  1 &   2 &   3 \\
b_1 & b_2 & b_3 
\end{array}
\right )
\;.
\eqlabel{2.9}\eeq
Note how we have applied the 
permutations on the left side
from right to left ($\leftarrow$). (Careful: Some authors apply them in
the opposite direction ($\rightarrow$)). A cycle is a special type of
permutation. If $G\in S_n$ maps 
$a_1\rarrow a_2$, $a_2\rarrow a_3$, \ldots, $a_{r-1}\rarrow a_r$,
$a_r\rarrow a_1$, where $i\neq j$ implies $a_i\neq a_j$ and $r\leq n$,
then we call $G$ a {\it cycle}. $G$ may be
 represented as in Eqs.(2.7) and (2.8).
Another way to represent it is by

\beq
G = ( a_1, a_2, a_3, \ldots, a_r)
\;.
\eqlabel{2.10}\eeq
(Careful: some people write $( a_r, \ldots, a_3, a_2, a_1)$ instead.)
We say that the cycle of Eq.(2.10) has {\it length} $r$. Cycles of length 1 are
just the identity map. A cycle of length 2 is called a {\it transposition}.
The product of two cycles need not be another cycle. For example,

\beq
(2,1,5) (1,4,5,6) =
\left ( 
\begin{array}{cccccc}
1 & 2 & 3 & 4 & 5 & 6\\
4 & 1 & 3 & 2 & 6 & 5 
\end{array}
\right )
\;
\eqlabel{2.11}\eeq
cannot be expressed as a single cycle. Any permutation
can be written as a product of cycles. For example,

\beq
\left ( 
\begin{array}{cccccc}
1 & 2 & 3 & 4 & 5 & 6\\
4 & 1 & 3 & 2 & 6 & 5 
\end{array}
\right )
=
(5,6) (1,4,2)
\;.
\eqlabel{2.12}\eeq
The cycles on the right side of Eq.(2.12) are {\it disjoint}; i.e., they
have no elements in common. Disjoint cycles commute. Any cycle  
can be expressed as a product of transpositions (assuming a group with $\geq 2$ elements),
 by using identities such as:

\beq
(a_1, a_2, \ldots, a_n) = 
(a_1, a_2) (a_2, a_3) \cdots (a_{n-1}, a_n)
\;,
\eqlabel{2.13}\eeq

\beq
(a_1, a_2, \ldots, a_n) = 
(a_1, a_n) \cdots (a_1, a_3)(a_1, a_2)
\;.
\eqlabel{2.14}\eeq
Another useful identity is

\beq
(a, b) = (a, p)(p, b)(a, p)
\;.
\eqlabel{2.15}\eeq
This last identity can be applied repeatedly. For example, applied 
twice, it gives

\beq
(a, b) = (a, p_1)(p_1, b)(a, p_1) = (a, p_1) (p_1, p_2) (p_2, b) (p_1, p_2) (a, p_1)
\;.
\eqlabel{2.16}\eeq
Since any permutation equals a product of cycles, and each 
of those cycles can be expressed as a product of transpositions,
all permutations can be 
expressed as a product of transpositions (assuming a group with $\geq 2$ elements). 
The decomposition of a 
permutation into transpositions is not unique. However,
the number of transpositions whose product equals a given
permutation is always either even or odd. An {\it even} (ditto, {\it odd}) {\it permutation} is
defined as one which equals an even (ditto, odd) number of transpositions.

\subsection*{2(d) Projection Operators}
\mbox{}\indent	
Consider a single qubit first.

The qubit's basis states $\ket{0}$ and $\ket{1}$ will
be represented by

\beq
\ket{0} =
\left[
\begin{array}{c}
1 \\ 0 
\end{array}
\right]
\;\;,\;\;
\ket{1} =
\left[
\begin{array}{c}
0 \\ 1 
\end{array}
\right]
\;.
\eqlabel{2.17}\eeq
The number operator $n$ of the qubit is defined by

\beq
n = 
\left[
\begin{array}{cc}
0 & 0 \\
0 & 1
\end{array}
\right]
=
\frac{ 1 - \sz}{2}
\;.
\eqlabel{2.18}\eeq
Note that 

\beq
n \ket{0} = 0
\;\;,\;\;
n \ket{1} = \ket{1}
\;.
\eqlabel{2.19}\eeq
We will often use $\antin$ as shorthand for 

\beq
\antin =
1-n = 
\left[
\begin{array}{cc}
1 & 0 \\
0 & 0
\end{array}
\right]
=
\frac{ 1 + \sz}{2}
\;.
\eqlabel{2.20}\eeq
Define $P_0$ and $P_1$ by

\beq
P_0 = \antin =
\left[
\begin{array}{cc}
1 & 0 \\
0 & 0
\end{array}
\right]
\;\;,\;\;
P_1 = n =
\left[
\begin{array}{cc}
0 & 0 \\
0 & 1
\end{array}
\right]
\;.
\eqlabel{2.21}\eeq
$P_0$ and $P_1$ are orthogonal projectors and they add to one:

\beq
P_a P_b = \delta(a, b) P_b
\;\;\;\;\; {\rm for} \;\; a,b\in Bool
\;,
\eqlabel{2.22a}\eeq

\beq
P_0 +  P_1 = I_2
\;.
\eqlabel{2.22b}\eeq

Now consider $\nb$ bits instead of just one. 

For 
$\beta\in \usualz$, we define $P_0(\beta)$, $P_1(\beta)$, $n(\beta)$
and $\antin(\beta)$ by means of Eq.(2.1).\enote{Havel}

For $\va \in Bool^\nb$, let

\beq
P_{\va} = P_{a_{\nb-1}} \otimes \cdots
\otimes P_{a_2} \otimes P_{a_1} \otimes P_{a_0}
\;.
\eqlabel{2.23}\eeq
Note that 

\beq
\sum_{\va\in Bool^\nb } 
P_{\va} =
I_2 \otimes I_2 \otimes \cdots \otimes I_2 = I_{2^\nb}
\;.
\eqlabel{2.24}\eeq
For example,
with 2 bits we have

\beq
P_{00} = P_0 \otimes P_0 = diag(1, 0, 0, 0)
\;,
\eqlabel{2.25a}\eeq

\beq
P_{01} = P_0 \otimes P_1 = diag(0, 1, 0, 0)
\;,
\eqlabel{2.25b}\eeq

\beq
P_{10} = P_1 \otimes P_0 = diag(0, 0, 1, 0)
\;,
\eqlabel{2.25c}\eeq

\beq
P_{11} = P_1 \otimes P_1 = diag(0, 0, 0, 1)
\;,
\eqlabel{2.25d}\eeq

\beq
\sum_{a,b \in Bool} P_{a,b} = I_4
\;.
\eqlabel{2.26}\eeq

For $r\geq 1$, suppose $P_1, P_2, \ldots P_r$ are orthogonal
projection operators (i.e., $P_i P_j = \delta(i, j) P_j$ ), and
$\alpha_1, \alpha_2 \ldots \alpha_r$ are complex numbers. Then it is easy
to show by Taylor expansion that 

\beq
\exp ( \sum_{i=1}^r \alpha_i P_i)
=
\sum_{i=1}^r \exp(\alpha_i) P_i +
( 1 - \sum_{i=1}^r P_i )
\;.
\eqlabel{2.27}\eeq
In other words, one can ``pull out" 
the sum over $P_i$'s from the argument of the exponential, but
only if one adds a compensating term $1 - \sum_i P_i$ so that 
both sides of the equation agree when all the $\alpha_i$'s are zero. 

\section*{3. State Permutations that Act on Two Bits}
\mbox{}\indent	
The goal of this paper is to reduce unitary matrices into 
qubit rotations and controlled-nots (c-nots). A qubit rotation
(i.e., $\exp[i\vec{\theta}\cdot\vec{\sigma}(\beta)]$
for $\beta\in \usualz$ and some real 3-dimensional vector $\vec{\theta}$) acts on 
a single qubit at a time. This section will discuss 
gates such as c-nots that are state permutations that  
act on two bits at a time.

\subsection*{3(a) $\nb = 2$}
\mbox{}\indent	
Consider first the case when there are only 2 bits. Then there
are four possible states--00, 01, 10, 11. With these 4 states,
one can build 6 distinct transpositions:

\beq
(00, 01) = 
\left [
\begin{array}{cc}
\sx 	&   \\
 		& I_2 
\end{array}
\right ]
=
P_0 \otimes \sx
+
P_1 \otimes I_2
=
\cnotno{1}{0}
\;,
\eqlabel{3.1a}\eeq

\beq
(00, 10) = 
\left [
\begin{array}{cc}
P_1 	& P_0  \\
P_0 	& P_1 
\end{array}
\right ]
=
I_2 \otimes P_1
+
\sx \otimes P_0
=
\cnotno{0}{1}
\;,
\eqlabel{3.1b}\eeq

\beq
(00, 11) = 
\left [
\begin{array}{ccc}
  &     & 1  \\
  & I_2 &    \\
 1&     &
\end{array}
\right ]
\;,
\eqlabel{3.1c}\eeq

\beq
(01, 10) = 
\left [
\begin{array}{ccc}
 1 &     &   \\
  & \sx &    \\
 &     & 1
\end{array}
\right ]
\;,
\eqlabel{3.1d}\eeq

\beq
(01, 11) = 
\left [
\begin{array}{cc}
P_0 	& P_1  \\
P_1 	& P_0 
\end{array}
\right ]
=
I_2 \otimes P_0
+
\sx \otimes P_1
=
\cnotyes{0}{1}
\;,
\eqlabel{3.1e}\eeq

\beq
(10, 11) = 
\left [
\begin{array}{cc}
I_2 	&   \\
 	    & \sx 
\end{array}
\right ]
=
 P_0 \otimes I_2 
+
 P_1 \otimes \sx
=
\cnotyes{1}{0}
\;,
\eqlabel{3.1f}\eeq
where matrix entries left blank should be interpreted as zero.
The rows and columns of the above matrices are labelled by 
binary numbers in increasing order (as in Eq.(2.3b) for $H_2$).
Note that the 4 transpositions Eqs.(3.1 a,b,e,f) change 
only one bit value, whereas the other 2 transpositions 
Eqs.(3.1 c,d) change both bit values. We will call 
$(00, 11)$ the {\it Twin-to-twin-er} and $(01, 10)$ the {\it Exchanger}.

Exchanger\enote{Fey} has four possible representations as
a product of c-nots:

\beq
(01, 10) = (01, 00) (00, 10) (01, 00) = 
\cnotno{1}{0}\cnotno{0}{1} \cnotno{1}{0}
\;,
\eqlabel{3.2a}\eeq

\beq
(01, 10) = (10, 11) (11, 01) (10, 11) = 
\cnotyes{1}{0}\cnotyes{0}{1} \cnotyes{1}{0}
\;,
\eqlabel{3.2b}\eeq

\beq
(01, 10) = (10, 00) (00, 01) (10, 00) = 
\cnotno{0}{1}\cnotno{1}{0} \cnotno{0}{1}
\;,
\eqlabel{3.2c}\eeq

\beq
(01, 10) = (01, 11) (11, 10) (01, 11) = 
\cnotyes{0}{1}\cnotyes{1}{0} \cnotyes{0}{1}
\;.
\eqlabel{3.2d}\eeq
Note that one can go 
from Eq.(3.2a) to (3.2b) 
by exchanging $n$ and $\antin$;
from Eq.(3.2a) to (3.2c)
by exchanging bit positions 0 and 1;
from Eq.(3.2a) to (3.2d) by doing both,
exchanging $n$ and $\antin$ and
exchanging bit positions 0 and 1. We will often represent Exchanger
by $E(0,1)$. It is easy to show that

\beq
E^T(0,1) = E(0,1) = E^{-1}(0, 1)
\;,
\eqlabel{3.3a}\eeq

\beq
E(0,1) = E(1,0)
\;,
\eqlabel{3.3b}\eeq

\beq
E^2(0,1) = 1
\;.
\eqlabel{3.3c}\eeq
Furthermore, if $X$ and $Y$ are any two arbitrary $2\times2$ matrices, 
then, by using the matrix representation Eq.(3.1d) of Exchanger,
one can show that 

\beq
E(1,0)\odot (X\otimes Y) = Y\otimes X
\;.
\eqlabel{3.4}\eeq
Thus, Exchanger exchanges the position of matrices $X$ and $Y$
in the tensor product.

Twin-to-twin-er also has 4 possible representations as a product of
c-nots. One is

\beq
(00, 11) = (00, 01) (01, 11) (00, 01) = 
\cnotno{1}{0}\cnotyes{0}{1} \cnotno{1}{0}
\;.
\eqlabel{3.5}\eeq
As with Exchanger, the other 3 representations are obtained by
replacing (1) $n$ and $\antin$, (2) bit positions 0 and 1, (3) both.

%don't float in general
%		\begin{figure}
		\begin{center}
			\epsfig{file=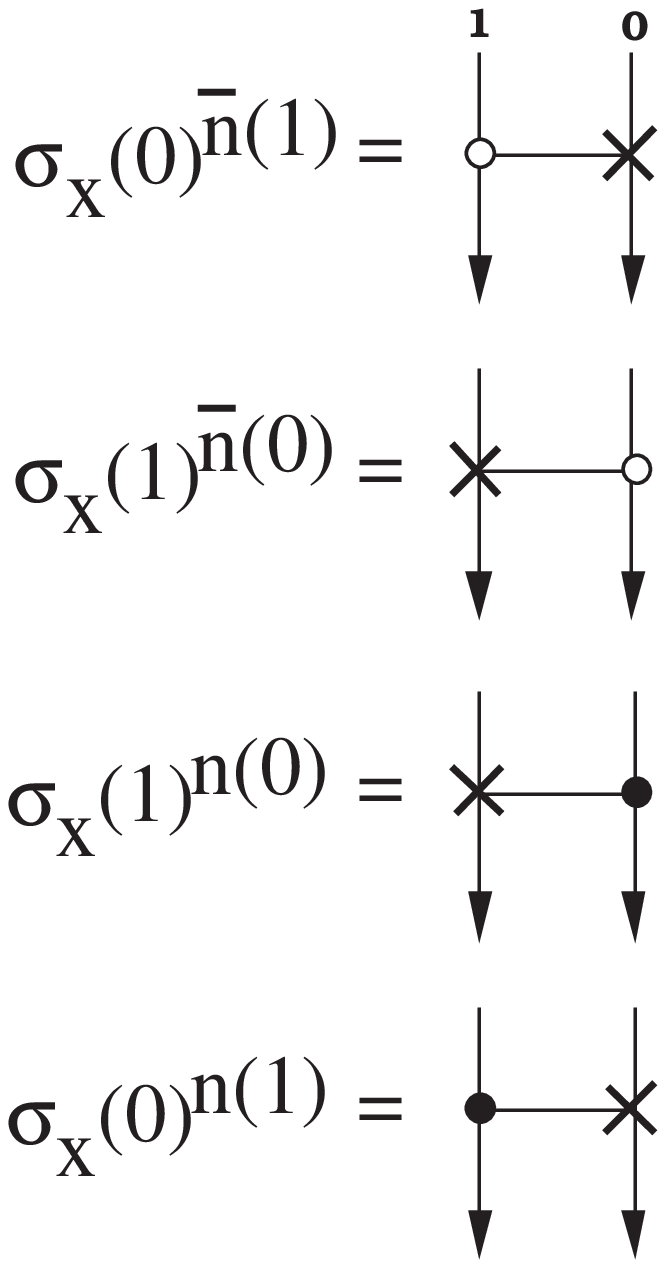}
			%\caption
			
			{Fig.2 Circuit symbols for the 4 different types of c-nots.}
			%\label{}
		\end{center}
%		\end{figure}

%don't float in general
%		\begin{figure}
		\begin{center}
			\epsfig{file=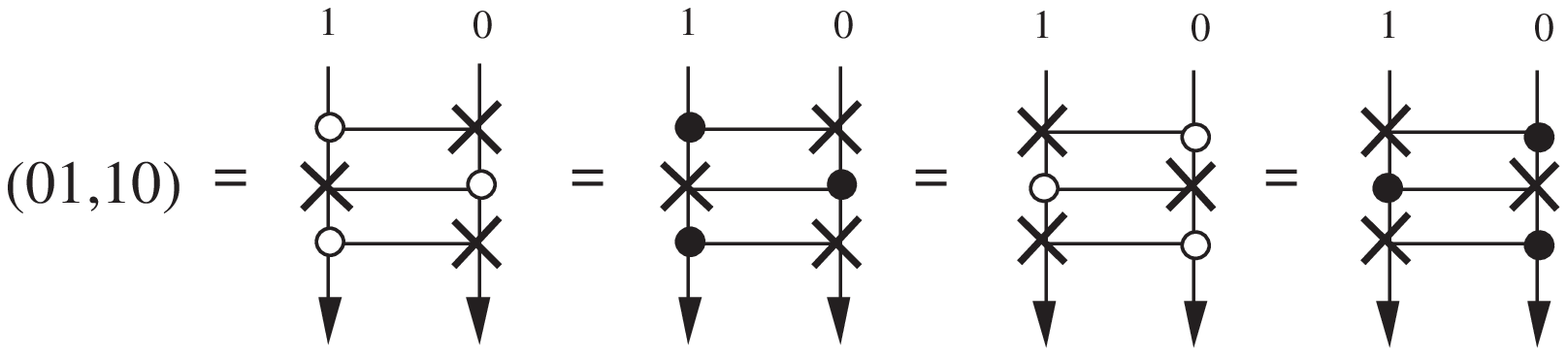}
			%\caption
			
			{Fig.3 Four equivalent circuit diagrams for Exchanger.}
			%\label{}
		\end{center}
%		\end{figure}

%don't float in general
%		\begin{figure}
		\begin{center}
			\epsfig{file=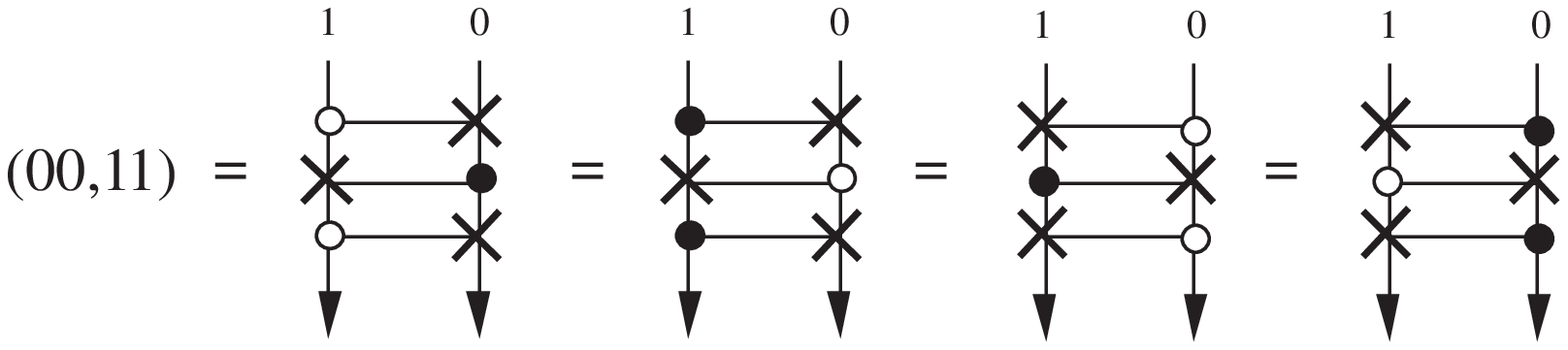}
			%\caption
			
			{Fig.4 Four equivalent circuit diagrams for Twin-to-twin-er.}
			%\label{}
		\end{center}
%		\end{figure}

%new paragraph after figure

Figures 2, 3 and 4 give a diagrammatic representation of the 6 
possible transpositions of states for $\nb=2$.

\subsection*{3(b) Any $\nb \geq 2$}
\mbox{}\indent
Suppose $a_1, b_1, a_2, b_2 \in Bool$ and
$\alpha, \beta \in \usualz$. We define 

\beq
(a_1 b_1, a_2 b_2)_{\alpha, \beta}
=
\prod_{ (\Lambda, \Lambda', \Lambda'')\in Bool^{\nb -2} }
(\Lambda a_1 \Lambda' b_1 \Lambda'', \Lambda a_2 \Lambda' b_2 \Lambda'')
\;,
\eqlabel{3.6}\eeq
where on the right side, $a_1, a_2$ are located at bit position $\alpha$,
and $b_1, b_2$ are located at bit position $\beta$. 
(Note that the transpositions
on the right side of Eq.(3.6) are disjoint so they commute.)
For example, for $\nb=3$, 

\beq
\cnotyes{1}{0} =
(10, 11)_{1,0} =
\prod_{a\in Bool}
(a10, a11) =
(010, 011) (110, 111)
\;.
\eqlabel{3.7}\eeq
Clearly, any permutation of states with $\nb$ bits
that permutes only 2 bits (i.e., Exchanger, Twin-to-twin-er, and all c-nots)
can be represented by
$(a_1 b_1, a_2 b_2)_{\alpha, \beta}$.

For $\alpha, \beta \in \usualz$, let $E(\alpha, \beta)$ represent
Exchanger:

\beq
E(\alpha, \beta) = (01, 10)_{\alpha, \beta}
\;.
\eqlabel{3.8}\eeq
As in the $\nb=2$ case, $E(\alpha, \beta)$ can be expressed as 
a product of c-nots in 4 different ways. One way is

\beq
E(\alpha, \beta) =
\cnotyes{\beta}{\alpha}
\cnotyes{\alpha}{\beta}
\cnotyes{\beta}{\alpha}
\;.
\eqlabel{3.9}\eeq
The other 3 ways are obtained by exchanging (1) $n$ and $\antin$,
(2) bit positions $\alpha$ and $\beta$, (3) both. Again as in the 
$\nb=2$ case,

\beq
E^T(\alpha, \beta) = E(\alpha, \beta) = E^{-1}(\alpha, \beta)
\;,
\eqlabel{3.10a}\eeq

\beq
E(\alpha, \beta) = E(\beta, \alpha)
\;,
\eqlabel{3.10b}\eeq

\beq
E^2(\alpha, \beta) = 1
\;.
\eqlabel{3.10c}\eeq
Furthermore, if $X$ and $Y$ are 
two arbitrary  $2\times 2$ matrices and 
$\alpha, \beta\in \usualz$ such that $\alpha\neq \beta$, then 

\beq
E(\alpha, \beta)\odot [X(\alpha) Y(\beta)]
=
X(\beta) Y(\alpha)
\;.
\eqlabel{3.11}\eeq

Equation (3.11) is an extremely useful result. It says that
$E(\alpha, \beta)$ is a transposition of bit positions. (Careful: this 
is not the same as a transposition of bit states.) 
Furthermore, the $E(\alpha, \beta)$ generate the group of $\nb !$ permutations of bit
positions. (Careful: this is not the same as the group of $(2^{\nb})!$
permutations of states with $\nb$ bits.)

%don't float in general
%		\begin{figure}
		\begin{center}
			\epsfig{file=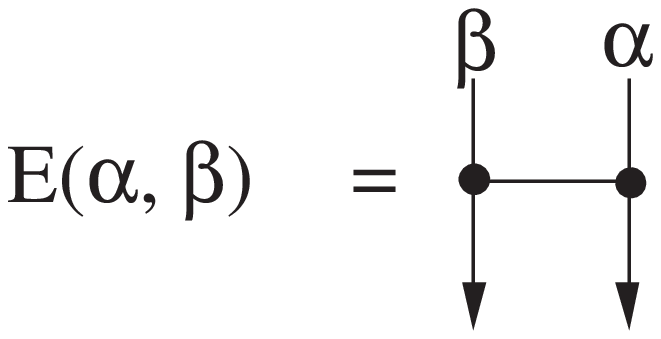}
			%\caption
			
			{Fig.5 Circuit symbol for Exchanger.}
			%\label{}
		\end{center}
%		\end{figure}

%don't float in general
%		\begin{figure}
		\begin{center}
			\epsfig{file=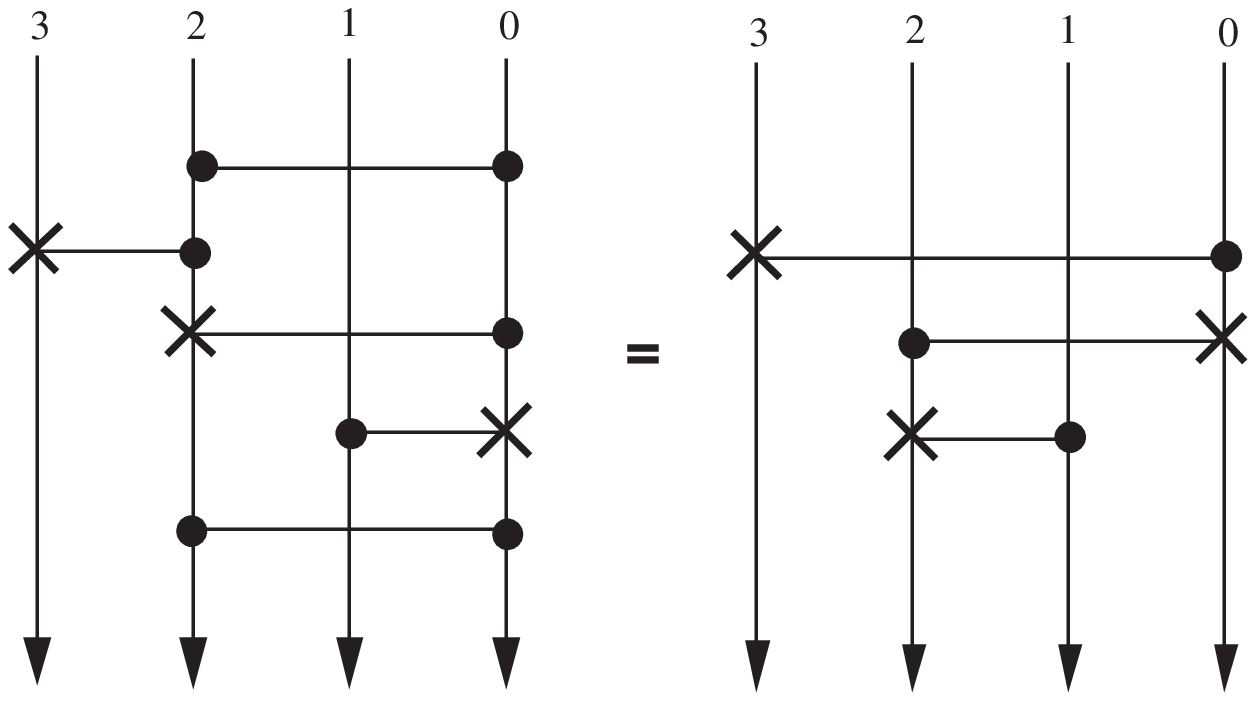}
			%\caption
			
			{Fig.6 Circuit diagram for Eq.(3.12).}
			%\label{}
		\end{center}
%		\end{figure}

%don't float in general
%		\begin{figure}
		\begin{center}
			\epsfig{file=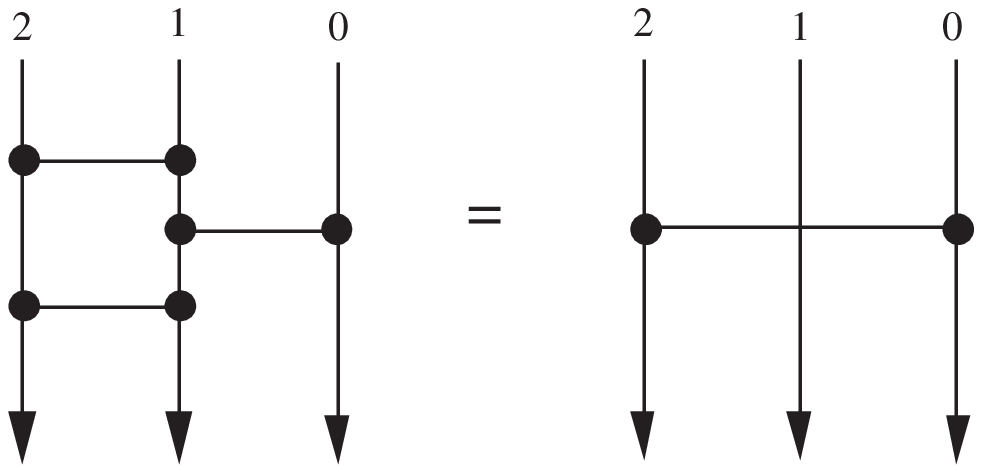}
			%\caption
			
			{Fig.7 Circuit diagram for Eq.(3.14).}
			%\label{}
		\end{center}
%		\end{figure}

%new paragraph after figures

An example of 
how one can use Eq.(3.11) is 

\beq
E(2, 0) \odot
[ \cnotyes{1}{0} \cnotyes{0}{2} \cnotyes{2}{3} ]
=
\cnotyes{1}{2} \cnotyes{2}{0} \cnotyes{0}{3}
\;.
\eqlabel{3.12}\eeq
Figure 5 gives a convenient way of representing $E(\alpha, \beta)$
diagrammatically. Using this symbol, the example of Eq.(3.12)
can be represented by Fig.6.

Of course, identities 
that are true for a general transposition are also true
for $E(\alpha, \beta)$. For example,

\beq
(2,0) = (2,1)(1,0)(2,1)
\;.
\eqlabel{3.13}\eeq
Therefore,
\beq
E(2,0) = E(2,1)E(1,0)E(2,1)
\;.
\eqlabel{3.14}\eeq
Figure 7 is a diagrammatic representation of Eq.(3.14).

\section*{4. Decomposing Central Matrix into SEO}
\mbox{}\indent	
In Section 1(d), we gave only a partial description of our algorithm.
In this section, we complete that description by showing how to
decompose each of the 3 possible kinds of central matrices
into a SEO.

\subsection*{4(a)When Central Matrix is a Single D Matrix}
\mbox{}\indent	
D matrices are defined by Eqs.(1.2). They can be expressed
in terms of projection operators as follows:

\beq
D = \sum_{\va \in Bool^{\nb-1}} \exp(i \phi_\va \sy )\otimes P_\va
\;,
\eqlabel{4.1}\eeq
where the $\phi_\va$ are real numbers. Note that in Eq.(4.1), $\va$ has
$\nb -1$ components instead of the full $\nb$. Using the identity Eq.(2.27),
one gets

\beq
D =
\exp
\left (
i \sum_{\va\in Bool^{\nb - 1} }
\phi_{\va} 
\sy \otimes P_\va
\right )
\;.
\eqlabel{4.2}\eeq
Now define new angles $\theta_\vb$ by

\beq
\phi_\va =
\sum_{\vb \in Bool^{\nb -1} }
(-1)^{\va\cdot\vb} \theta_\vb
\;.
\eqlabel{4.3}\eeq
Suppose $\vec{\phi}$ (ditto, $\vec{\theta}$) is a column vector
whose components are the numbers $\phi_\va$ (ditto, $\theta_\va$)
arranged in order of increasing $\va$. Then Eq.(4.3) is equivalent
to 

\beq
\vec{\phi} = H_{\nb -1} \vec{\theta}
\;.
\eqlabel{4.4}\eeq
This is easily inverted to 

\beq
\vec{\theta} = 
\frac{1}{ 2^{\nb -1} }
H_{\nb - 1} \vec{\phi}
\;.
\eqlabel{4.5}\eeq
Let $A_\vb$ for $\vb \in Bool^{\nb - 1}$ be defined by

\beq
A_\vb =
\exp
\left (
i \theta_\vb 
\sy
\otimes
\sum_{\va \in Bool^{\nb-1}}
(-1)^{\va\cdot\vb}
P_\va
\right )
\;.
\eqlabel{4.6}\eeq
Then $D$ can be written as

\beq
D =
\prod_{ \vb \in Bool^{\nb -1} }
A_\vb
\;.
\eqlabel{4.7}\eeq
Note that the $A_\vb$ operators on the right side commute
so the order in which they are multiplied is irrelevant.
Next we establish 2 useful identities:
If $\beta\in Z_{0, \nb -1} $ and $\vec{u}(\beta) \in Bool^{\nb }$
is the $\beta$'th standard unit vector, then 

\beq
\begin{array}{l}
\sum_{\va \in Bool^{\nb } }
(-1)^{\va \cdot \vec{u}(\beta) }
P_\va \\
= 
I_2 \otimes
\cdots\otimes
I_2\otimes
\left [\sum_{a_\beta \in Bool}
(-1)^{a_\beta}
P_{a_\beta}
\right ]
\otimes
I_2 \otimes
\cdots\otimes
I_2\\
=
P_0(\beta) - P_1(\beta)\\
=
\sz(\beta)
\end{array}
\;.
\eqlabel{4.8}\eeq
If $\beta, \alpha \in \usualz $ and $\alpha \neq \beta$, then

\beq
\begin{array}{l}
\cnotyes{\alpha}{\beta}
\odot
\sy(\beta)\\
=
[ \sx(\beta) P_1(\alpha) + P_0(\alpha) ]
\odot
\sy(\beta)\\
=
\sy(\beta) 
[ - P_1(\alpha) + P_0(\alpha) ]\\
=
\sy(\beta) \sz(\alpha)
\end{array}
\;.
\eqlabel{4.9}\eeq
Now we are ready to express $A_\vb$
in terms of elementary operators. For any $\vb \in Bool^{\nb -1} $, 
we can write

\beq
\vb = \sum_{j=0}^{r-1} \vec{u}(\beta_j)
\;,
\eqlabel{4.10}\eeq
where 

\beq
\nb-2 \geq 
\beta_{r-1} 
> \cdots
> \beta_1 > \beta_0 \geq 0
\;.
\eqlabel{4.11}\eeq
In other words, $\vb$ has bit value of 1 at bit positions 
$\beta_j$. At all other bit positions, $\vb$ has
bit value of 0. $r$ is the number of bits in $\vb$ whose
value is 1. When $\vb =0$, $r=0$. Applying Eqs.(4.8) and (4.9), one
gets

\beq
\begin{array}{l}
[
\cnotyes{\beta_{r-1}}{\nb -1 }
\cdots
\cnotyes{\beta_2}{\nb -1 }
\cnotyes{\beta_1}{\nb -1 }
]
\odot
\sy (\nb -1)\\
=
\sy(\nb -1) 
\prod_{j=0}^{r-1}
\sz(\beta_j)\\
=
\prod_{j=0}^{r-1}
\left(
\sum_{\va }
(-1)^{\va\cdot\vec{u}(\beta_j)}
\sy \otimes P_\va
\right )\\
=
\sum_{\va }
(-1)^{\va \cdot \vb}
\sy \otimes P_\va\
\end{array}
\;.
\eqlabel{4.12}\eeq
Thus,

\beq
A_\vb
= 
[
\cnotyes{\beta_{r-1}}{\nb -1 }
\cdots
\cnotyes{\beta_2}{\nb -1 }
\cnotyes{\beta_1}{\nb -1 }
]
\odot
\exp[ i\theta_\vb \sy(\nb -1) ]
\;.
\eqlabel{4.13}\eeq
There are other ways of decomposing $A_\vb$ into a SEO. For example,
using the above method, one can also show that

\beq
A_\vb
= 
[
\cnotyes{\beta_{r-1}}{\beta_{r-2} }
\cdots
\cnotyes{\beta_2}{\beta_1}
\cnotyes{\beta_1}{\beta_0}
]
\odot
\exp[ i\theta_\vb \sy(\beta_0) ]
\;.
\eqlabel{4.14}\eeq

In conclusion, we have shown how to decompose a D matrix into
a SEO. For example, suppose $\nb = 3$. Then

\beq
D = \sum_{a,b \in Bool}
\exp ( i \phi_{ab} \sy ) \otimes
P_a \otimes P_b
\;.
\eqlabel{4.15}\eeq
Define $\vec{\theta}$ by 

\beq
\vec{\theta} = \frac{1}{4} H_2 \vec{\phi}
\;.
\eqlabel{4.16}\eeq
By Eqs.(4.7) and (4.13),

\beq
D = A_{00} A_{01} A_{10} A_{11}
\;,
\eqlabel{4.17}\eeq
where

\beq
A_{00} = \exp( i \theta_{00} \sy )
\otimes I_2 \otimes I_2
\;,
\eqlabel{4.18a}\eeq

\beq
A_{01} = 
\cnotyes{0}{2}\odot
[\exp( i \theta_{01} \sy )
\otimes I_2 \otimes I_2 ]
\;,
\eqlabel{4.18b}\eeq

\beq
A_{10} = 
\cnotyes{1}{2}\odot
[\exp( i \theta_{10} \sy )
\otimes I_2 \otimes I_2 ]
\;,
\eqlabel{4.18c}\eeq

\beq
A_{11} = 
[\cnotyes{1}{2}
\cnotyes{0}{2}]
\odot
[\exp( i \theta_{11} \sy )
\otimes I_2 \otimes I_2 ]
\;.
\eqlabel{4.18d}\eeq

\subsection*{4(b)When Central Matrix is a Direct Sum of D Matrices}
\mbox{}\indent
Consider first the case $\nb = 3$. Let $R(\phi) = \exp(i \sy \phi) $.
Previously we used the fact that any D matrix $D$ can be
expressed as 

\beq
D = \sum_{a, b\in Bool}
R(\phi_{ab}'') \otimes P_a \otimes P_b
\;.
\eqlabel{4.19}\eeq
But what if $R$ were located at bit
positions 0 or 1 instead of 2? By expressing both sides of
the following equations as $8 \times 8$ matrices, one can show that

\beq
D_0 \oplus D_1
=
\sum_{a,b\in Bool}
P_a \otimes R(\phi_{ab}') \otimes P_b
\;,
\eqlabel{4.20}\eeq

\beq
D_{00}\oplus D_{01}\oplus D_{10}\oplus D_{11}
=
\sum_{a,b\in Bool}
P_a \otimes P_b \otimes R(\phi_{ab})
\;,
\eqlabel{4.21}\eeq
where the $D_j$ and $D_{ij}$ are D matrices.
One can apply a string of Exchangers to move $R$ 
in Eqs.(4.20) and (4.21) to any bit position. Thus,

\beq
D_0 \oplus D_1
=
E(1,2) \odot
\left (
\sum_{a,b\in Bool}
 R(\phi_{ab}') \otimes P_a \otimes P_b
\right )
\;,
\eqlabel{4.22}\eeq

\beq
D_{00}\oplus D_{01}\oplus D_{10}\oplus D_{11}
=
E(0,1) E(1,2) \odot
\left (
\sum_{a,b\in Bool}
R(\phi_{ab}) \otimes 
P_a \otimes  
P_b
\right )
\;.
\eqlabel{4.23}\eeq
(Careful: $E(0, 2) \neq E(0,1) E(1,2)$. $E(0,2)$ will change
$\sum_{a,b} R(\phi_{ab})\otimes P_a \otimes P_b$
to 
$\sum_{a,b} P_b \otimes P_a \otimes R(\phi_{ab})$, which 
is not the same as the left side of Eq.(4.21) ).

For general $\nb \geq 1$, 
if $k\in \usualz$ and 

\beq
E =
\left \{
\begin{array}{l}
1 \;\;\;\; {\rm if} \;\; k=0 \;\; {\rm or} \;\; \nb=1\\
E(\nb - k - 1, \nb -k)\cdots
E(\nb - 3, \nb - 2)
E(\nb - 2, \nb - 1)
\;\;\;\; {\rm otherwise}
\end{array}
\right .
\;,
\eqlabel{4.24}\eeq
then 
a direct sum of $2^k$  D matrices
can be expressed as 

\beq
E\odot
\left (
\sum_{\va\in Bool^{\nb -1}} 
R(\phi_{\va})
\otimes P_\va
\right )
\;.
\eqlabel{4.25}\eeq

It follows that if we want to decompose a direct sum of D matrices
into a SEO,
we can do so in 2 steps: (1) decompose 
into a SEO the D matrix that one obtains
by moving the qubit rotation to bit position $\nb -1$, (2)
Replace each bit name in the decomposition by its ``alias".
By alias we mean the new name assigned by the 
bit permutation $E$ defined by Eq.(4.24).

\subsection*{4(c)When Central Matrix is a Diagonal Unitary Matrix}
\mbox{}\indent
Any diagonal unitary matrix $C$ can be expressed as

\beq
C = \sum_{\va \in Bool^{\nb}} \exp(i \phi_\va)  P_\va
\;,
\eqlabel{4.26}\eeq
where the $\phi_\va$ are real numbers. Using the identity Eq.(2.27) yields

\beq
C =
\exp
\left (
i \sum_{ \va \in Bool^\nb }
\phi_\va
P_\va
\right )
\;.
\eqlabel{4.27}\eeq
Now define new angles $\theta_\vb$ by

\beq
\phi_\va =
\sum_{\vb \in Bool^{\nb} }
(-1)^{\va\cdot\vb} \theta_\vb
\;.
\eqlabel{4.28}\eeq
In terms of vectors, 

\beq
\vec{\phi} = H_{\nb} \vec{\theta}
\;,
\eqlabel{4.29}\eeq
and

\beq
\vec{\theta} = 
\frac{1}{ 2^{\nb} }
H_{\nb} \vec{\phi}
\;.
\eqlabel{4.30}\eeq
Let $A_\vb$ for $\vb \in Bool^\nb$ be defined by

\beq
A_\vb =
\exp
\left(
i \theta_\vb 
\sum_{\va \in Bool^{\nb}}
(-1)^{\va\cdot\vb}
P_\va
\right)
\;.
\eqlabel{4.31}\eeq
Then $C$ can be written as

\beq
C =
\prod_{ \vb \in Bool^{\nb} }
A_\vb
\;,
\eqlabel{4.32}\eeq
where the $A_\vb$ operators commute. For any $\vb \in Bool^\nb$,
we can write

\beq
\vb = \sum_{j=0}^{r-1} \vec{u}(\beta_j)
\;,
\eqlabel{4.33}\eeq
where 

\beq
\nb-1 \geq 
\beta_{r-1} 
> \cdots
> \beta_1 > \beta_0 \geq 0
\;.
\eqlabel{4.34}\eeq
(Careful: Compare this with Eq.(4.11). Now
$\vb\in Bool^\nb$ instead of $Bool^{\nb -1}$ and
$\beta_{r-1}$ can be as large as $\nb-1$ instead of $\nb-2$.)
One can show using the techniques of Section 4(a) that

\beq
A_\vb
=
\left \{
\begin{array}{l}
\exp[i \theta_0] 
\;\;\;\; {\rm if} \;\;\; r=0 
\\
\exp[i \theta_\vb \sz(\beta_0)] 
\;\;\;\; {\rm if} \;\;\; r=1 
\\
\left [\cnotyes{\beta_{r-1}}{\beta_0 }
\cdots
\cnotyes{\beta_2}{\beta_0}
\cnotyes{\beta_1}{\beta_0}
\right ]\odot
\exp[ i\theta_\vb \sz(\beta_0) ]
\;\;\;\;{\rm if}\;\;\; r\geq 2
\end{array}
\right . 
\;.
\eqlabel{4.35}\eeq
As in Section 4(a), there are other ways of decomposing $A_\vb$ into a SEO.

In conclusion, we have shown how to decompose a 
diagonal unitary matrix into a SEO. For example, suppose $\nb = 2$. Then

\beq
C = diag( e^{i \phi_{00}}, e^{i \phi_{01}}, e^{i \phi_{10}},e^{i \phi_{11}} )
\;.
\eqlabel{4.36}\eeq
Define $\vec{\theta}$ by 

\beq
\vec{\theta} = \frac{1}{4} H_2 \vec{\phi}
\;.
\eqlabel{4.37}\eeq
By Eqs.(4.32) and (4.35),

\beq
C = A_{00} A_{01} A_{10} A_{11}
\;,
\eqlabel{4.38}\eeq
where

\beq
A_{00} = \exp( i \theta_{00} )
\;,
\eqlabel{4.39a}\eeq

\beq
A_{01} =
I_2 \otimes
\exp( i \theta_{01} \sz )
\;,
\eqlabel{4.39b}\eeq

\beq
A_{10} = 
\exp( i \theta_{10} \sz )
\otimes I_2 
\;,
\eqlabel{4.39c}\eeq

\beq
A_{11} = 
\cnotyes{1}{0}
\odot
[
I_2 \otimes
\exp( i \theta_{11} \sz )
]
\;.
\eqlabel{4.39d}\eeq

\section*{5. Qubiter}
\mbox{}\indent	
At present, Qubiter is a very 
rudimentary program. We hope that its fans will extend and enhance it in the future.
Qubiter1.0  is  written in pure C++, and has no graphical user interface.
In its ``compiling" mode, Qubiter takes as input
a file with the entries of a unitary matrix and returns as output
a file with a SEO. In its ``decompiling"
mode, it does the opposite: it takes a SEO file and returns the entries of a matrix.
The lines in a SEO file are of 
4 types:

\begin{description}

\item{(a) PHAS \qquad $ang$}\newline 
where $ang$ is a real number. This signifies 
a phase factor $\exp(i(ang)\frac{\pi}{180})$.

\item{(b) CNOT \qquad $\alpha$ \qquad $char$ \qquad $\beta$}\newline 
where $\alpha, \beta\in \usualz$ and $char\in \{ T, F \}$.
$T$ and $F$ stand for true and false.
If $char$ is the character $T$, this signifies $\cnotyes{\alpha}{\beta}$.
Read it as ``c-not: if $\alpha$ is true, then flip $\beta$."
If $char$ is the character $F$, this signifies $\cnotno{\alpha}{\beta}$.
Read it as ``c-not: if $\alpha$ is false, then flip $\beta$."

\item{(c) ROTY \qquad $\alpha$ \qquad $ang$}\newline 
where $\alpha \in \usualz$ and $ang$ is a real number. This
signifies the rotation of qubit $\alpha$ about the Y axis by an angle $ang$ in degrees.
In other words, $\exp(i\sy(\alpha)ang\frac{\pi}{180})$.

\item{(d) ROTZ \qquad $\alpha$ \qquad $ang$}\newline 
This is the same as (c) except that the rotation is about
the Z axis instead of the Y one.

\end{description} 

As a example, consider what Qubiter gives for 
the Discrete Fourier Transform matrix $U$. This matrix
has entries

\beq
U_{\va, \vb} =
\frac{1}{\sqrt{\ns}}
\exp
\left[
\frac{i 2\pi d(\va) d(\vb) }{\ns}
\right ]
\;,
\eqlabel{5.1}\eeq
where $\va, \vb \in Bool^{\nb}$. $\ns$ and $d(\cdot)$ were
defined in Section 2(a). When $\nb = 2$,
Qubiter gives 33 operations (see Fig. 8).
After doing the trivial optimization of removing all 
factors $A_\vb$ for which the rotation angle is zero,
the 33 operations in Fig.8 reduce to 25 operations in Fig.9. In the
future, we plan to introduce into Qubiter many more 
optimizations and some quantum error correction. 
Much work remains to be done.

%don't float in general
%		\begin{figure}
		\begin{center}
			\epsfig{file=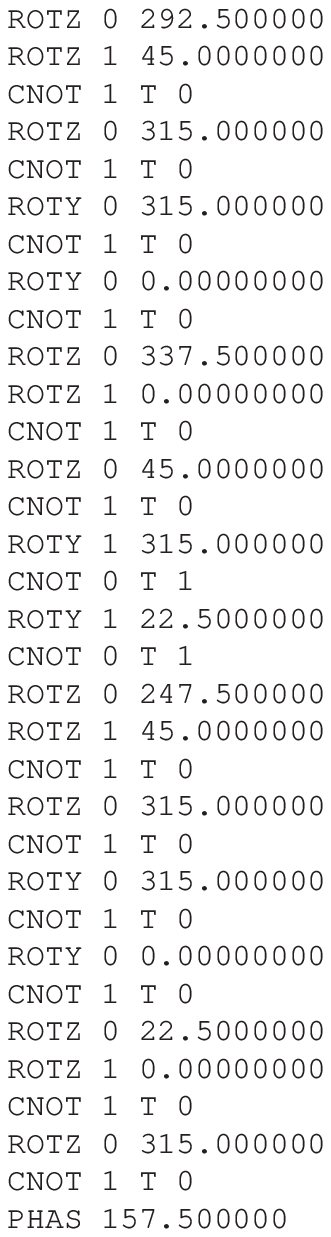}
			%\caption
			
			{Fig.8 Output of Qubiter with
			zero angle optimization turned OFF.
			(input: 2-bit Discrete Fourier Transform matrix). }
			%\label{}
		\end{center}
%		\end{figure}

%don't float in general
%		\begin{figure}
		\begin{center}
			\epsfig{file=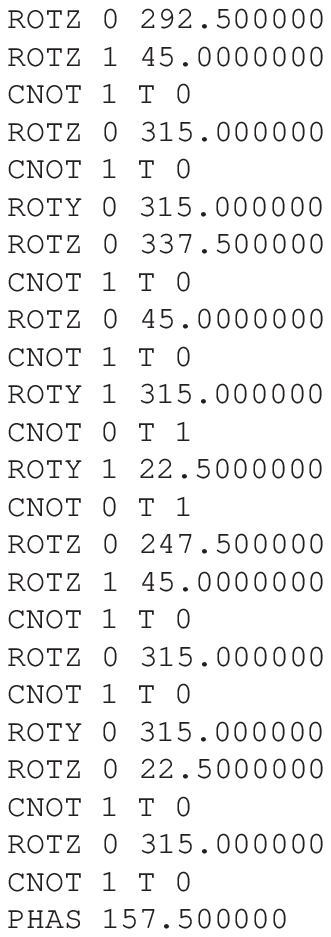}
			%\caption
			
			{Fig.9 Output of Qubiter with
			zero angle optimization turned ON.
			(input: 2-bit Discrete Fourier Transform matrix). }
			%\label{}
		\end{center}
%		\end{figure}

\newpage
\section*{FIGURE CAPTIONS:}
\begin{description}

\item{\sc Fig.1} A binary tree. Each node
$\beta$ has a single parent. If the parent is to 
$\beta$'s right (ditto, left), then $\beta$
contains the names of the matrices produced by
applying the CS Decomposition Theorem to the $L$
matrices (ditto, 
$R$ matrices) of $\beta$'s parent.

\item{\sc Fig.2} Circuit symbols for the 4 different types of c-nots.

\item{\sc Fig.3} Four equivalent circuit diagrams for Exchanger.

\item{\sc Fig.4}  Four equivalent circuit diagrams for Twin-to-twin-er.

\item{\sc Fig.5} Circuit symbol for Exchanger.

\item{\sc Fig.6} Circuit diagram for Eq.(3.12).

\item{\sc Fig.7} Circuit diagram for Eq.(3.14).

\item{\sc Fig.8} Output of Qubiter with
	zero angle optimization turned OFF.
	(input: 2-bit Discrete Fourier Transform matrix).
	
\item{\sc Fig.9} Output of Qubiter with
	zero angle optimization turned ON.
	(input: 2-bit Discrete Fourier Transform matrix).
	
\end{description}

\end{document}